\documentclass[traditabstract]{aa} % for the abstract without structuration 
\usepackage{graphicx}
\usepackage{txfonts}
\usepackage{natbib}
\bibpunct{(}{)}{;}{a}{,}

\newcommand\lesim{\lower.5ex\hbox{$\; \buildrel < \over \sim \;$}}
\newcommand\gesim{\lower.5ex\hbox{$\; \buildrel > \over \sim \;$}}
\newcommand\HI{H{\small I}}
\newcommand\kms{km s$^{-1}$}

\newcommand\He{He}
\newcommand\Htwo{H$_2$}
\newcommand\msun{M$_{\odot }$}
\newcommand\msunpccube{M$_{\odot }$pc$^{-3}$}

\newcommand\lsun{L$_{\odot }$}
\newcommand\magarcsecsq{mag arcsec$^{-2}$}

\begin{document}
   \title{The dark matter halo shape of edge-on disk galaxies}

\subtitle{IV. UGC 7321}

\titlerunning{The dark matter halo shape of edge-on disk galaxies IV}
   \author{J.C. O'Brien
          \inst{1},
K.C. Freeman\inst{1}
          \and
          P.C. van der Kruit\inst{2}}

   \institute{Research School of Astronomy and Astrophysics, Australian National
University, Mount Stromlo Observatory, Cotter Road, ACT 2611, Australia\\
              \email{jesscobrien@gmail.com; kcf@mso.anu.edu.au$^\star$}
         \and
             Kapteyn Astronomical Inst[Bitute, University of Groningen, P.O. Box 800,
9700 AV Groningen, the Netherlands\\
             \email{vdkruit@astro.rug.nl}\thanks{For 
correspondence contact Ken Freeman or Piet van der Kruit}
             }
\authorrunning{J.C. O'Brien et al.}

   \date{Received Xxxxxxx 00, 2010; accepted Xxxxxxx 00, 2010}

% \abstract{}{}{}{}{} 
% 5 {} token are mandatory
 
  \abstract
   {This is the fourth paper in a series in which we attempt to put
constraints on the flattening of dark halos in disk galaxies. We
observed for this purpose the \HI\ in edge-on galaxies, where it is in principle
possible to measure the force field in the halo vertically and radially
from gas layer flaring and rotation curve decomposition respectively.  
As reported in earlier papers in this series we have for
this purpose analysed the \HI\ channel maps to accurately measure all four 
functions that describe as a function of galactocentric radius the planar \HI\ 
kinematics and 3D \HI\ distribution of a galaxy: the radial \HI\ surface density, 
the \HI\ vertical thickness, the rotation curve and the \HI\ velocity dispersion. 
In this paper we analyse these data for the edge-on galaxy UGC7321. 

We measured the stellar mass distribution ($M=3\times10^8$ \msun 
with $M/L_R\lesim0.2$), finding that the vertical force of 
the gas disk dominates the 
stellar disk at all radii. Measurements of both the rotation curve and the
vertical force field showed that the vertical force puts a much
stronger constraint on the stellar mass-to-light ratio than rotation 
curve decomposition. Fitting of the vertical force field derived from the 
flaring of the \HI\ layer and \HI\ velocity dispersion revealed that UGC7321 
has a spherical halo density distribution with a flattening of 
$q = c/a = 1.0 \pm 0.1$. However, 
the shape of the vertical force field showed that a non-singular isothermal 
halo was required, 
assuming a vertically isothermal \HI\ velocity dispersion. 
A pseudo-isothermal halo and a 
gaseous disk with a declining \HI\ velocity dispersion at high latitudes may 
also fit the vertical 
force field of UGC7321, but to date there is no observational evidence 
that the \HI\ velocity dispersion declines away from the galactic plane. 
We compare the halo flattening of UGC7321 with other 
studies in the literature and discuss its implications. 
Our result is consistent with new n-body simulations which show that 
inclusion of hydrodynamical modelling produces more spherical halos.
}
   {}


   \maketitle
%
%________________________________________________________________

\section{Introduction}

In paper I in this series \citep{ofk2008a} we presented \HI\
observations of a sample of 8 edge-on, \HI\ rich, late-type galaxies.
The aim of the project has been described there in detail. Briefly,
we attempt to put constraints on the flattening of dark halos around
disk galaxies by measuring the force field of the halo vertically
from the flaring of the \HI\ layer and radially from rotation curve
decomposition. For the vertical force field we need to determine
in these galaxies both the velocity dispersion of the \HI\ gas
(preferably as a function of height from the central plane of the
disk) and the thickness of the \HI\ layer, all of this as a function
of galactocentric radius. In addition we also need to extract
information on the rotation of the galaxy and the deprojected
\HI\ surface density, also as a function of galactocentric radius.

In paper II \citep{ofk2008b} we discussed methods to analyse the \HI\
observations in edge-on galaxies and presented a new method to 
measure the radial distributions, rotation curves and velocity
dispersions. We applied this method to our sample of 
galaxies in the third paper in this series: \citet{ofk2008c}. In that paper
we also developed a new method to derive the thickness of the \HI\ layer, 
or `flaring profile', 
as a function of galactocentric radius, which we used to measure the
\HI\ flaring of each galaxy in our sample. 

In the present paper we have fitted the vertical shape $q=c/a$ of the halo
density distribution for the northern galaxy UGC7321. This particular system 
was chosen as a first application since 
the sensitivity of the \HI\ imaging obtained for UGC7321 at
the VLA was 5 times greater than that for the southern galaxies that we
observed with the ATCA, allowing more accurate measurement of the
gas layer flaring to high latitudes, and better measurement of the \HI\
velocity dispersion. Due to its northern location, UGC7321 was not in
our initial southern galaxy sample for which we measured near-IR and
optical stellar photometry at Siding Spring Observatory. 
However, VLA \HI\ data observed by Lyn Matthews \citep{mgvd1999} 
was available for this galaxy and Michael Pohlen \citep{pdla2002} kindly
supplied $R$-band photometry which allowed us to derive the
stellar luminosity density necessary to analyse the halo density
distribution.

In Sect.~\ref{sec:ch7-stars} we present the surface brightness
and deprojected luminosity volume density, and our derivation of the halo
core radius, halo asymptotic velocity, and stellar mass-to-light ratio
by rotation curve decomposition. In Sect.~\ref{sec:ch7-halo_method}, we
present a new simple method used to measure the halo shape using the
vertical gradient of the vertical force, $dK_z/dz$, with the
usual assumption of gas in hydrostatic equilibrium to determine the total
$dK_z/dz$ of the galaxy. The resulting halo shape measurement
for UGC7321 is presented and discussed in Sect.~\ref{sec:ch7-halo_fit}, 
and compared to other measurements of dark halo flattening in
Sect.~\ref{sec:ch7-comparison}. Sect.~\ref{sec:ch7-summ} summarizes our
conclusions.

\section{Stellar surface brightness and deprojected luminosity density of
  UGC7321} 
\label{sec:ch7-stars}

\begin{figure}[t]
  \centering
\includegraphics[width=9cm]{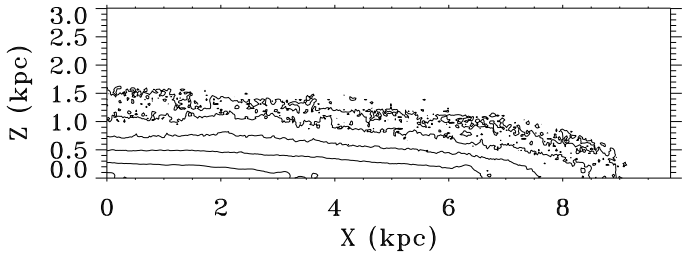} 
  \caption[UGC7321: Projected stellar surface density averaged over
  all quadrants.]{Projected stellar surface density of UGC7321
    averaged over all quadrants. Contours are separated on a log scale
    at 0.1, 0.3, 1, 3, 10, 30 L$_{\odot}$pc$^{-2}$. }
  \label{fig:ch7-lsun_pc_sq}
\end{figure}

Fig.~\ref{fig:ch7-lsun_pc_sq} shows the $R$-band surface brightness 
-- averaged over four quadrants -- of
UGC7321 with contours ranging from 0.1 to 30 L$_{\odot}$pc$^{-2}$ in
steps of 0.5 dex.  The observations and photometric calibration of
UGC7321 are discussed in \citet{pdla2002}. \citet{pbld2003} analysed
the projected surface brightness showing peanut-shaped deviations from
elliptical fits to the isophotes at $z$ heights greater than 0.5 kpc
above the plane. These deviations provide strong evidence of a
stellar bar, although it is difficult to measure the scale of the
bar from the scale of the boxy-peanut shaped bulge (Athanassoula,
private communication).

%%% Figure %%%
% Luminosity surface density

Using an exponential radial surface profile we fit the central surface
brightness and apparent radial scale length of the projected surface
density.  In logarithmic units
\begin{equation}
\mu(X) = \mu_0 + 1.086(X/h_X) \ \ {\rm mag \, arcsec}^{-2},
\end{equation}
where $X$ denotes the major axis distance.  We found the 
projected central $R$-band surface brightness to be 22.0 \magarcsecsq,
with a scalelength of $4.0\pm0.3$ kpc, in agreement with that measured
by \citet{pbld2003}. The projected central surface brightness is
consistent with the $B$ and $R$-band measurement by \citet{mgvd1999} when
the internal extinction model that \citeauthor{mgvd1999} used is
factored in.  The projected surface brightness shows a small nuclear
feature smaller than 1 kpc, and an exponential profile from 0.5 to
6.5 kpc, declining steeply after 7 kpc.

\subsection{Deprojection}
\label{ch7:stars_deproj}

To deproject the luminosity distribution from the edge-on projection,
we assume azimuthal symmetry and perform a direct deprojection of the
projected surface density on the sky using the inverse Abel transform
\begin{equation}
\label{ch7:abel}
I(R,z) = \frac{-1}{\pi} \int_R^\infty \frac{1}{ \sqrt{X^2 - R^2} }
\frac{d I(X,z)}{dX} dX,
\end{equation}
where $X$ is the position along the major axis and $R$ is the
galactocentric radius in the cylindrical coordinate system.  Applying
the inverse Abel tranform to each $z$-plane yields the luminous volume
density $I(R,z)$ of the galaxy in L$_{\odot}$pc$^{-3}$ as a function of
cylindrical radius $R$ and height $z$. To minimise the incidence of
local regions with a negative derivative $d I(X,z)/dX$, we
performed the deprojection after smoothing the projected surface
density by several different 2D Gaussians with FWHM ranging from
256 pc to 1027 pc (or 10 and 40 pixels, respectively).

%%% Figure %%%
% Luminosity volume density
\begin{figure}[t]
 \centering
\includegraphics[width=9cm]{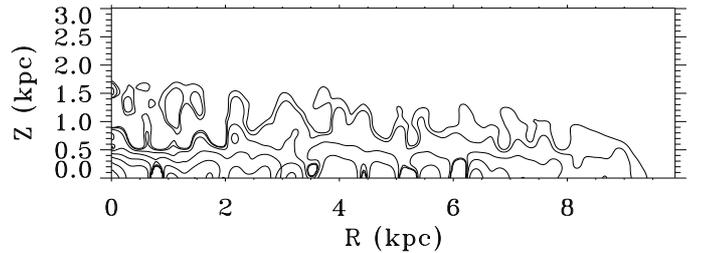} 
  \caption[UGC7321: Stellar luminosity volume density as a function of $R$ and
  $z$ after deprojecting via the Inverse Abel Integral]
  {Stellar luminosity volume density in $R$ and $z$ after deprojecting the
    observed surface density using the Inverse Abel Integral. Contours
    are separated on a log scale at 0.0001, 0.003, 0.001, 0.003, 0.01,
    0.03, 0.1, 0.3 L$_{\odot}$pc$^{-3}$. The radial and vertical scalelengths,
    as shown in Fig.~\ref{fig:ch7-faceon_sersic_fits}, are
    $h_R=2.65$ kpc and $h_z=245$ pc.}
  \label{fig:ch7-lsun_pc_cube}
\end{figure}

\begin{figure}[t]
  \centering 
\includegraphics[width=9cm]{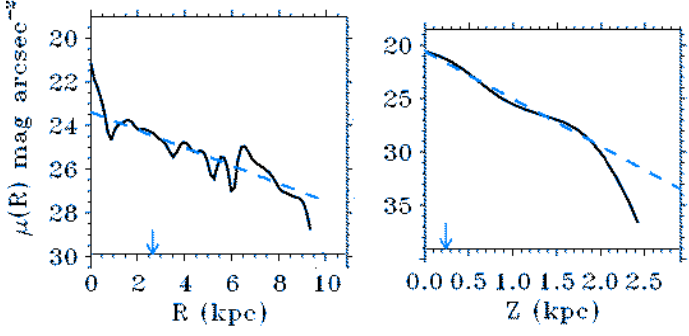}
  \caption[UGC7321: Exponential fits to the radial and vertical stellar surface
  brightness profiles formed by integrating over the deprojected
  luminosity volume density]{Exponential fits to the radial and
    vertical surface brightness profiles formed by integrating over
    the deprojected volume density. The arrows mark the radial
    scalelength (left) and vertical scaleheight (right).}
  \label{fig:ch7-faceon_sersic_fits}
\end{figure}

Fig.~\ref{fig:ch7-lsun_pc_cube} shows the derived volume density
$I(R,z)$ of UGC7321, while Fig.~\ref{fig:ch7-faceon_sersic_fits}
shows the radial and vertical surface brightness obtained by
integrating over the volume density. UGC7321 is indeed a very low
surface brightness galaxy. The face-on radial scalelength is $h_R =
2.65\pm0.17$ kpc. Both the
volume density and the radial profile, show that UGC7321 clearly has a
small central nuclear region that is approximately 8 times brighter
in surface brightness than the fitted central surface brightness of
$\mu_R=23.4\pm0.14$ \magarcsecsq\ derived assuming an exponential disk
only. The peak face-on central surface brightness of the nuclear
region is $\mu_R=21.1$ \magarcsecsq.

The central luminosity volume density of UGC7321 is 0.3 L$_{\odot}$pc$^{-3}$
in the central nucleus averaged over 250 pc. At 2.2 scalelengths the
midplane volume density is 0.002 L$_{\odot}$pc$^{-3}$, much less than the
luminosity density near the Sun ($\approx 0.1$ L$_{\odot}$pc$^{-2}$) as would
be expected for such a low surface brightness galaxy.

\subsection{Rotation curve decomposition}
\label{sec:ch7-rotcurve_decomp}

In the standard manner we decomposed the rotation curve to obtain the
parameters of a spherical pseudo-isothermal halo, and constrain the
stellar mass-to-light ratio. The radial surface density of the stars
and gas was used directly to derive the rotation curve contribution
due to each luminous mass component, and the observed rotation curve
fitted such that
\begin{equation}
v_h^2 = v_{obs}^2 - \left( v_s^2 + v_g^2 \right),
\end{equation}
where $v_{obs}(R)$ is the observed rotation curve, and $v_h(R)$,
$v_s(R)$ and $v_g(R)$ are the rotation due to the halo, stellar and
gas mass components, respectively. The radial surface density of each
luminous component was obtained by integrating over $z$, 
and a constant $R$-band stellar $M/L_R$ was adopted. The \HI\
surface density was scaled by a factor of 1.4 to include \He\ and
(a minimal amount of) molecular gas. 

The observed rotation curve was fitted using a spherical
pseudo-isothermal halo density distribution
\begin{equation}
\rho_h(R) = \rho_{h,0} \frac{R_c^2}{R^2 + R_c^2} 
\end{equation}
with corresponding rotation curve
\begin{equation}
v_h^2(R) = 4 \pi G \rho_{h,0} R_c^2 \left[ 1 -
  \frac{R_c}{R} \mathrm{arctan}\left(\frac{R}{R_c}\right) \right].
\end{equation}
 The pseudo-isothermal
halo shown above is defined by the core radius $R_c$ and central density
$\rho_{h,0}$, and asymptotes at $R\rightarrow\infty$ to 
\begin{equation}
v_{h,\infty}^2 = 4\pi G \rho_{h,0} R_c^2,
\end{equation}
such that the rotation can also be written
\begin{equation}
v_h^2(R) = v_{h,\infty}^2 \left[ 1 -
  \frac{R_c}{R} \mathrm{arctan}\left(\frac{R}{R_c}\right) \right].
\end{equation}

By definition the rotation curve measures the total force in the
radial direction $K_R = v^2(R)/R$; however it is unable to
constrain the halo flattening $q$. 

By evaluating the radial force
$K_R(R,z)$ of a flattened pseudo-isothermal halo with density distribution
\begin{equation}
\rho(R,z) = \frac{\rho_{h,0}R_c^2}{R_c^2 + R^2 + z^2/q^2},
\end{equation}
\citet{srjf1994} show that the corresponding rotation curve 
in the midplane is
\begin{eqnarray}
  v_h^2(R,q)&=& v_{h,\infty}^2(q) \left[ 1 -  {\gamma \over {\arctan \gamma}} 
\left( {{ q^2 R_c^2}\over{R^2 + a}} \right)^{1/2}
\right.  \nonumber \\
&&\left.
    \arctan \left( {R^2 + a}\over {q^2 R_c^2} \right)^{1/2} 
\right],
\end{eqnarray}
where 
\begin{equation}
  \gamma = \frac{ \sqrt{1 - q^2} }{q} \ \ \ ;\ \ \ 
   a = (1 - q^2)R_c^2 \nonumber
\end{equation}
and the asymptotic rotation $v_{h,\infty}^2(q)$ is very similar to
the asymptotic rotation of a spherical isothermal halo with 
\begin{equation}
\label{eq:ch7-v_inf_q}
  v_{h,\infty}^2(q) = 4\pi G\rho_{h,0} R_c^2 f(q),
\end{equation}
where 
\begin{equation}
  f(q) = \frac{ q  }{ \sqrt{1 - q^2} } \mathrm{arccos}(q).
\end{equation}

%%% Figure %%%
% Face-on sersic fit

We use the rotation curve of a spherical pseudo-isothermal halo, as
the shape of the rotation of a similar, but flattened halo is
almost the same. This similarity implies that the measured asymptotic
rotation derived from a spherical pseudo-isothermal fit to the
rotation curve also defines the asymptotic rotation of a flattened
pseudo-isothermal halo via Eqn.~(\ref{eq:ch7-v_inf_q}).

\begin{figure}[t]
  \centering
\includegraphics[width=9cm]{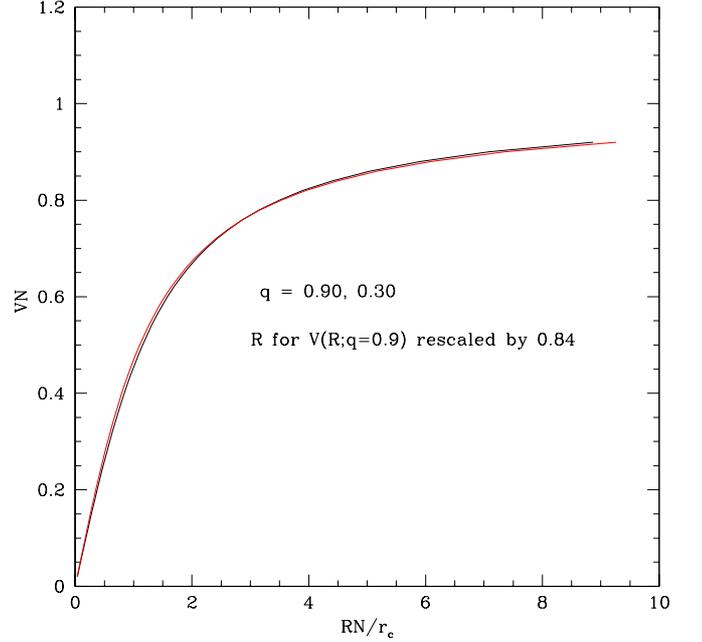}
  \caption[Scaled rotation curves of a flattening pseudo-isothermal
  halo for two different values of the halo flattening $q$]{Scaled
    rotation curves of a flattening pseudo-isothermal halo for two
    different values of the halo flattening $q$. The thick, black curve shows
    the rotation curve of $q=0.3$ and the thin, red curve shows the rotation
    for $q=0.9$ with the radius scaled by 0.84.  The vertical axis
    shows the rotation normalised by the asymptotic rotation
    $v_{h,\infty}(q)$, while the abscissa shows the radius normalised
    by the core radius $R_c$. The nearly identical shape of the two
    curves shows that shape of the rotation curve of the halo is
    almost independent of $q$, with the radial scaling varying by 
a constant factor (of only
    $\approx15\%$ in this case) over a large range of $q$.}
  \label{fig:ch7-rc_indpdt}
\end{figure}

In Fig.~\ref{fig:ch7-rc_indpdt}  we
show how the rotation curve of a flattened pseudo-isothermal halo varies
with $q$ (for $q\leq1$) at z=0. The vertical axis shows the rotation
normalised by the asymptotic rotation $v_{h,\infty}(q)$, while the
abscissa shows the radius normalised by the core radius $R_c$. The thick,
black curve shows the rotation for $q=0.3$, while the thin, red curve shows
the rotation for $q=0.9$ with the radius scaled by 0.84. The nearly
identical shape of the two curves shows that shape of the rotation
curve of the halo is almost independent of $q$, with the radial
scaling varying by only $\approx15\%$ over a large range of $q$. The
derived halo central density does vary significantly with $q$,
becoming denser for more flattened halos:
\begin{equation}
\rho_{h,0}(q) = \frac{ v_{h,\infty}^2(q=1) }{4 \pi \, G}
\frac{1}{R_c^2 \, f(q)},
\end{equation}

%%% Figure %%%
% v_h as fn of q

In Sect.~\ref{sec:ch7-halo_fit} we measure $q$ by fitting the
gradient of the vertical force $dK_z/dz$, where the gradient of the 
flattened pseudo-isothermal halo is also given by \citet{srjf1994}. As
$K_z(R,z)$ also depends on the same parameters $q$, halo core
radius $R_c$ and halo central density $\rho_{h,0}$, we can use the
asymptotic halo rotation $v_{h,\infty}$ and the core radius $R_c$
which are well measured from rotation curve decomposition, and just
perform a single parameter fit to the flattening $q$, as the central
density $\rho_{h,0}$ is tied to $v_{h,\infty}$, $R_c$ and $q$.

\begin{figure}[t]
 \centering
\includegraphics[width=9cm]{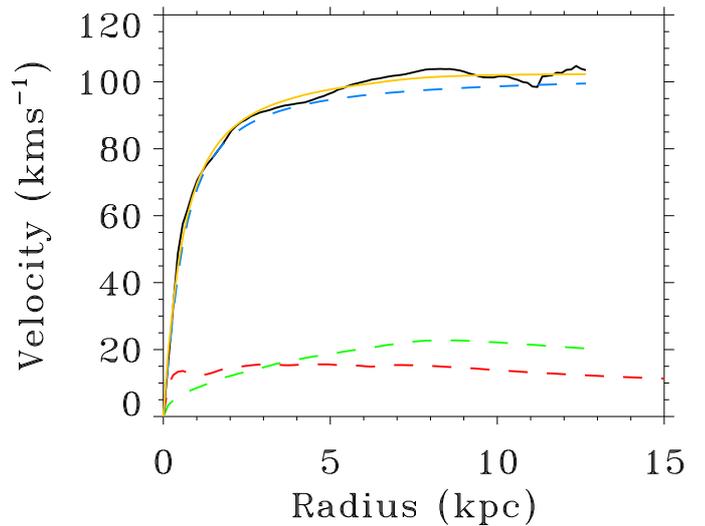}
  \caption[UGC7321: Rotation curve decomposition]{Rotation curve
    decomposition of UGC7321. The observed rotation curve is shown by
the thick  (black) line, while the rotation due to the other components
are shown by dashed lines; from the bottom up stars (red), gas (green) and 
the halo (blue).  The disk has in this fit an $M/L_R$ of 1.05.
The resulting fit,
    $v_{fit}^2 = v_g^2 + v_s^2 + v_h^2$, is shown by the grey full-drawn
line (yellow).}.
  \label{fig:ch7-rc_fit}
\end{figure}

In Fig.~\ref{fig:ch7-rc_fit} we show the best fitting stellar
and halo rotation curves derived from rotation curve decomposition of
UGC7321.  The observed rotation curve is shown by the thick black line, 
while the rotation due to gas, stars and the halo are shown by dashed 
lines (from the bottom up stars (red), gas (green) and halo (blue). 
The resulting fit is shown by the grey full-drawn line (yellow). The best fit
was achieved with a sub-maximal (see below) stellar $M/L_R$ of 1.05, which
scaled the stellar component from a luminosity of $L=4.0\times10^{8}$
\lsun\ to a mass of $4.2\times10^8$ \msun. UGC7321 is a gas-rich low
surface brightness galaxy with a $M_{HI}/L_R=2.2$; thus even with an
$M/L_R$ near unity, the stellar mass is roughly a third of the gas mass
($M_g = 1.4 M_{HI}$).

 The plot shows that the rotation curve is well
fit by a pseudo-isothermal halo with core radius $R_c=0.52\pm0.02$
kpc, and central density $\rho_{h,0}=0.73\pm0.05$ \msunpccube. 
As the halo core radius and halo asymptotic rotation are nearly
independent of the halo shape in flattened pseudo-isothermal halos, we
adopt those measurements from rotation curve decomposition
for our halo shape modelling in Sect.~\ref{sec:ch7-halo_method}. The
halo is clearly the dominant component at all radii, comprising over
90\%\ of the total galaxy mass of $3.2\times10^{10}$ \msun\ at the
last measured point. \citet{um2003} reach a very similar conclusion. 
UGC7321 has a similar dark-to-light mass ratio, although it is 
significantly more massive than DDO154 and shows evidence of a bar
\citep{cf1988}.

The dominance of the halo at all radii means that the stellar
mass-to-light ratio is rather unconstrained.  Forcing a maximum
disk fit yielded a stellar $M/L_R$ of 2.9, but resulted in a very
poor fit with rotation velocities in error by 15-25\%\ in the inner disk. Even with
 a maximum disk fit, the peak rotation for the stellar disk was only 43
\kms. The best fit occurs for an
$M/L_R$ of 1.05, but all $M/L_R$ values less than 1.05 provide acceptable 
fits to the rotation curve. Consequently, we use 1.05 as an upper limit in 
the analysis in Sect.~\ref{sec:ch7-halo_method}. 

The low mass-to-light ratio and extremely sub-maximal nature of the UGC7321
stellar disk implies that the recent star formation rate exceeds the average 
rate over the galaxy's lifetime. Observations by \citet{mgvd1999} that found a 
significant fraction of young stars support our results. Detailed
studies of the vertical disk structure indicate multiple disk components \citep{mat2000}.
Our finding of a very low mass-to-light ratio 
(in the $R$-band) warrants further studies of the structure
and composition of the stellar disk, but this is beyond the scope of the 
present paper.

%%% Figure %%%\includegraphics[]{../../../../../Desktop/2001ASPC__230__187F.pdf}

% RC fit

%%% Figure %%%
% Unfitted gradient - stellar M/L = 1.05
\begin{figure*}[t]
  \centering 
\includegraphics[width=15cm]{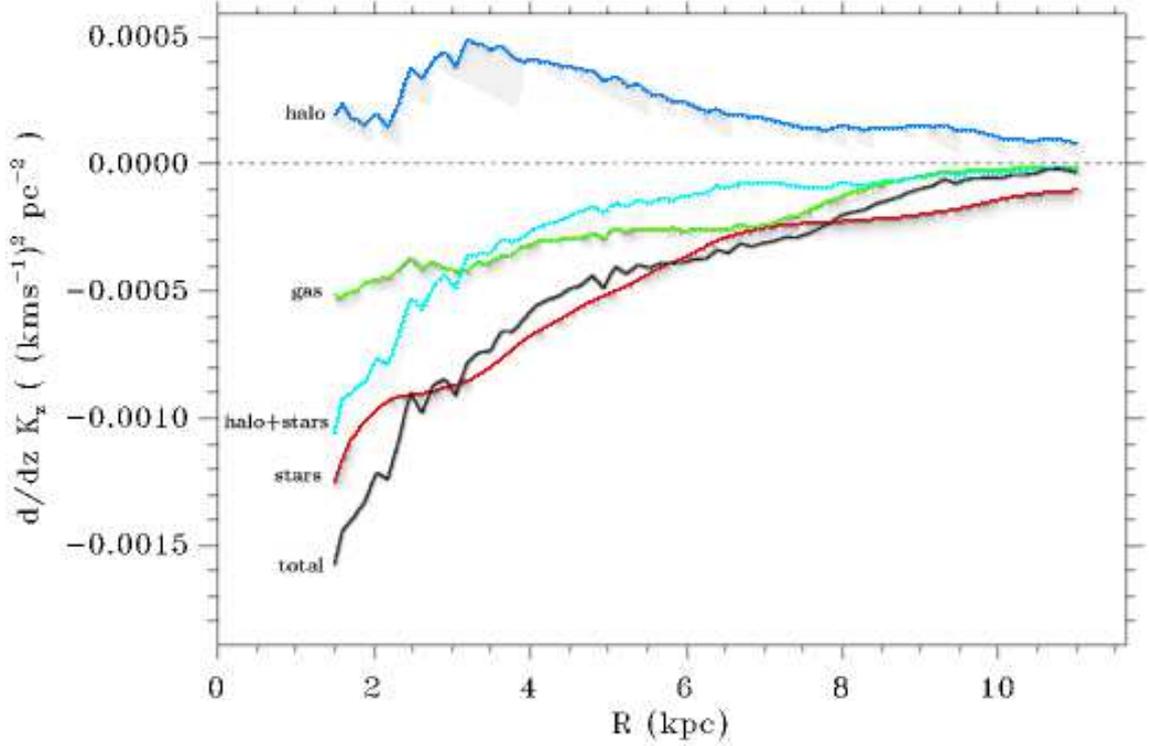} 
  \caption[UGC7321: The $z$-gradient of the total vertical force, showing
  total, gas, stellar and halo components, where the stellar component
  uses the stellar $M/L_R$ measured from rotation curve
  decomposition]{The vertical gradient of the vertical force. The
    total gradient, $dK_{z,tot}/dz$ determined from hydrostatics is
    shown as the thick, black line (labelled `total'), 
while the gradients for the gas and 
stars is shown in green and red (and labelled `gas' and `stars'), 
respectively. The stellar gradient $dK_{z,s}/dz$
    is calculated using the stellar mass-to-light ratio of
    $M/L_R=1.05\pm0.33$, derived from rotation curve decomposition. This
    value of the stellar $M/L_R$ is clearly too high as the gradient
    allowed for the halo (shown in blue and labelled `halo') 
is positive at all radii, requiring negative halo densities. The
    stellar $M/L_R$ needs to be reduced to allow a negative halo
    gradient $dK_{z,h}/dz$. The cyan line (labelled `halo+stars')
shows the sum of stellar and
    halo components determined by $dK_{z,tot}/dz - dK_{z,g}/dz$, which
    must be fitted with a non-negligible stellar $M/L_R$ and a positive
    density halo.}
  \label{fig:ch7-grad_mleq1}
\end{figure*}

\section{Method for fitting the halo shape}
\label{sec:ch7-halo_method}

The thickness of the gas layer depends on the vertical force $K_z$,
and hence on the shape of the dark matter halo
\citep{maloney1992,ovg1992,khg1993,maloney1993}. A gas layer with
less flaring for a given gas velocity dispersion implies a stronger
$K_z$.

Analytically it has been shown that at large radii the thickness of the gas
layer is roughly proportional to the square root of the halo
flattening $q$ \citep{maloney1993,olling1995} and that the flaring
should increase radially in an exponential fashion
\citep{vdkruit1981}. This was confirmed in our measurements and
the earlier study of  \citet{rupen1991}
comprising high resolution VLA \HI\ observations of NGC891 and NGC4565.

We determine the halo flattening by measuring the $z$-gradient of the
total vertical force $K_{z,tot}(R,z)$ from the equation of hydrostatic
equilibrium for the gas layer, and evaluating $dK_z/dz$ for each
luminous mass component using Poisson's equation. The halo gradient
$dK_{z,h}/dz$ is modelled using the equation for the vertical force
$K_z$ of a flattened pseudo-isothermal halo given in \citet{srjf1994}.

Given the gas disk is in equilibrium, the gas pressure
gradient and internal forces must exactly balance the gradient of the 
total gravitational potential of the galaxy, 
where the total gravitational potential $\Phi_{tot}$ is the sum 
of the stellar, gas and halo potentials, $\Phi_{tot} =
\Phi_h + \Phi_s + \Phi_g$.  Assuming that the gas velocity dispersion 
is isothermal in $z$ (though
not in $R$), the equilibrium condition in $z$ becomes 
\begin{equation}
  \frac{\partial ^2}{\partial z^2} [\sigma_{v,g}^2(R) \, \mathrm{ln} \rho_g(R,z)]
    =   \frac{\partial K_z(R,z)}{\partial z}.
\end{equation}

If we further assume that the gas density distribution is
Gaussian in $z$, the vertical gradient of the total $K_z$ becomes
a simple function of the gas velocity dispersion and the vertical
FWHM thickness of the gas, both functions of radius which we
measured in paper III, namely
\begin{equation}
\frac{\partial K_{z,tot}(R,z)}{\partial z} = - 
  \frac{\sigma_{v,g}^2(R) }{ ({\rm FWHM}_{z,g}(R)/2.355)^2}.
\label{eqn:total_gradient}
\end{equation}
From this we see that the vertical gradient of $K_{z,tot}$ derived
from such a gas disk is constant in $z$, and varying in $R$. 

The gradient of the vertical force of each of the luminous components
was directly calculated from the Poisson equation for each component
\begin{equation}
\label{eq:ch7-poisson-1}
\frac{\partial K_z(R,z)}{\partial z} =
 - 4\pi G \rho(R,z) + \frac{1}{R}\frac{\partial (R K_R)}{\partial R},
\end{equation}
where $\rho(R,z)$, $K_R$, and $K_z$ correspond to the volume density 
and forces of that component.
In terms of $v^2(R)$, the squared rotation due to that mass component,
we can rewrite this as
\begin{equation}
\label{eq:ch7-poisson-2}
\frac{\partial K_z(R,z)}{\partial z} =
 -4\pi G \rho(R,z) + \frac{1}{R} \frac{\partial v^2(R)}{\partial R}.
\end{equation}

Consequently, the gradient of the halo force must satisfy
\begin{eqnarray}
\frac{\partial K_{z,h}(R,z)}{\partial z} &=& \frac{\partial
  K_{z,tot}(R,z)}{\partial z} - \frac{\partial
  K_{z,s}(R,z)}{\partial z} - \nonumber \\
 && \frac{\partial K_{z,g}(R,z)}{\partial z},
\end{eqnarray}
where the subscripts $s$, $g$ and $h$, denote the stars, gas and halo,
respectively.

The vertical $\partial K_{z,h}(R,z)/\partial z$ force of the flattened
pseudo-isothermal halo is given in \citet{srjf1994}. As
the asymptotic halo rotation and the halo core radius were well
determined from the rotation curve decomposition, and the central
density determined by $q$, the fitting of $dK_{z,h}/dz$ reduces to a
fit with a single parameter $q$. 

Comparison of the stellar vertical force gradient with the total
vertical force gradient shows that the vertical force puts a much
stronger constraint on the stellar mass-to-light ratio than does the
radial force fitting undertaken in rotation curve decomposition.
Inclusion of the gas self-gravity requires that the stellar vertical
force gradient $dK_{z,s}(R,z)/dz$ must be
\begin{equation}
\frac{d K_{z,s}(z)}{dz} < \frac{d K_{z,tot}(z)}{dz} - \frac{d
  K_{z,g}(z)}{dz}
\end{equation}
where from here on we take the $R$-dependence of $K_z$ as implicit and
write the derivatives as total derivatives.

Given that the stellar mass density and its squared rotation are both
proportional to the stellar mass-to-light ratio, we see from
Eqn.~(\ref{eq:ch7-poisson-2}) that the vertical gradient of the vertical
force of the stars is linearly related to the stellar mass-to-light
ratio.  As the low stellar luminosity meant the rotation curve
decomposition was relatively insensitive to the stellar mass-to-light
ratio, we consider the stellar mass-to-light ratio to be a free
parameter and fitted
$$
\frac{dK_{z,tot}}{dz} - \frac{dK_{z,g}}{dz}
$$
 as determined from the observations as a function of $R$ with  
$$
\frac{dK_{z,h}}{dz} + \frac{dK_{z,s}}{dz},
$$
where the $K_{z,h}$ is modelled by the flattened
pseudo-isothermal halo and the $z$-gradient of the stellar vertical
force is 
$$
\frac{dK_{z,s}}{dz} =
\left.\left(\frac{M}{L}\right)\frac{dK_{z,s}}{dz}\right| _{M/L_R=1}.
$$ 
The total
$K_z$ gradient and the gas $K_z$ gradient come directly from
observations without any free parameters. 

We recall from above that the vertical gradient of $K_{z,tot}$
derived for a Gaussian gas disk is independent of $z$.  To be most
sensitive to the constraints from the luminous mass density, we fit
the gradient of the vertical force near the midplane at $z=100$ pc,
high enough to avoid the bulk of the internal extinction caused by
dust in the plane of the thin stellar disk.

\section{The halo shape of UGC7321}
\label{sec:ch7-halo_fit}

\subsection{Results of the fitting}

Fig.~\ref{fig:ch7-grad_mleq1} shows the vertical gradient of each of
the $K_z$ components for UGC7321. The thick, black line (labelled `total')
is the gradient determined from the gas flaring and velocity distribution
using Eqn.~(\ref{eqn:total_gradient}). The gradient due to the stellar disk 
is shown by the red line (labelled `stars') and has been calculated
with the mass-to-light ratio of 1.05, which was the best fitting
value in our rotation curve decomposition. We see immediately
that the stellar $dK_{z,s}/dz$ alone is comparable to the total
$dK_{z,tot}/dz$ given by the hydrostatics over most of the range of
$R$. The gradient due to the \HI\ is the line labelled `gas' (green). 
We can subtract this gradient for the gas from the total gradient and
derive the gradient due to the sum of the halo and the stars. This is the
(cyan) line labelled `halo+stars'. Subtracting the gradient for the stars
from the gradient for the (halo + stars) then leaves the gradient that 
should be attributed to the halo alone (blue line labelled `halo'). 
With the adopted M/L ratio for the stellar disk, the halo gradient turns
out to be positive, which is unphysical.

From this example, it is clear that the
gradient of $K_{z,tot}$ measured from the hydrostatics provides a very
strong constraint on the stellar $M/L_R$ ratio. Even with a zero-mass
halo, which is excluded by the rotation curve fit, we see 
that the stellar $M/L_R$ must be less than 1 to leave room for the
gradient of $K_z$ given by the gas self-gravity.  With
the necessary inclusion of the gas, we find that the stellar
mass-to-light ratio $M/L_R$ must be $\lesim 0.6$.  The difference
$dK_{z,tot}/dz - dK_g/dz$ (cyan; labelled `halo+stars') 
constitutes the combined $dK_z/dz$ of the halo (blue; labelled `halo') 
and stars (red; labelled `stars').  The rotation curve decomposition
requires a positive halo mass density at all radii, thus $dK_{z,h}/dz
< 0$ for all radii. This constraint requires that the stellar $M/L_R \ll
1$.

%%% Figure %%%
% Form of dK_{z,h}/dz - for varying q
\begin{figure}[t]
  \centering
\includegraphics[width=9cm]{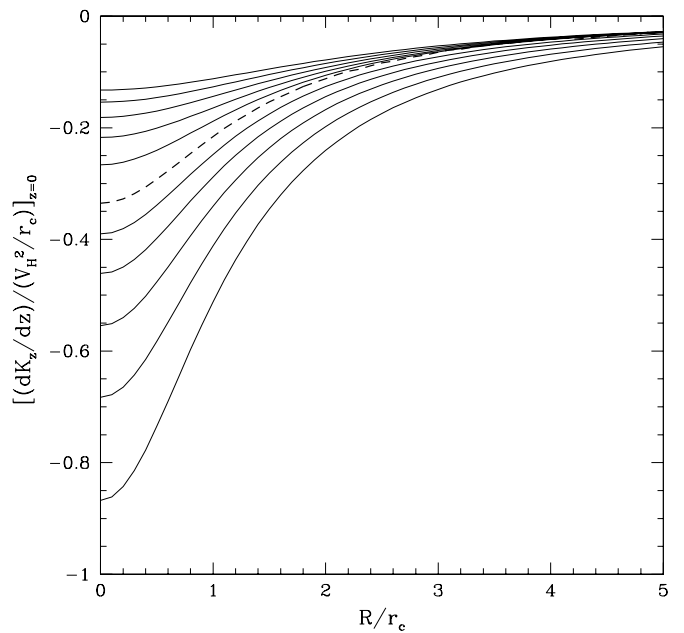} 
  \caption[The vertical gradient equated near the midplane of a
  flattened pseudo-isothermal halo for $q$ from 0.5 to 1.5]{The
    vertical gradient equated near the midplane of a flattened
    pseudo-isothermal halo, as calculated from Eqn.~(6) of
    \citet{srjf1994} for both oblate and prolate halos ranging from
    $q$ of 0.5 to 1.5 in steps of 0.1. The bottom curve
    corresponds to the most flattened halo with $q=0.5$, becoming
    increasingly shallow as the halo gets less flat. The gradient of a
    spherical halo is shown by the dashed line. The top curve
    corresponds to $q=1.5$.}
  \label{fig:ch7-dKdz_halo_vary_q}
\end{figure}

%%% Figure %%%
% IT & PIT halos
\begin{figure}[t]
 \centering
\includegraphics[width=9cm]{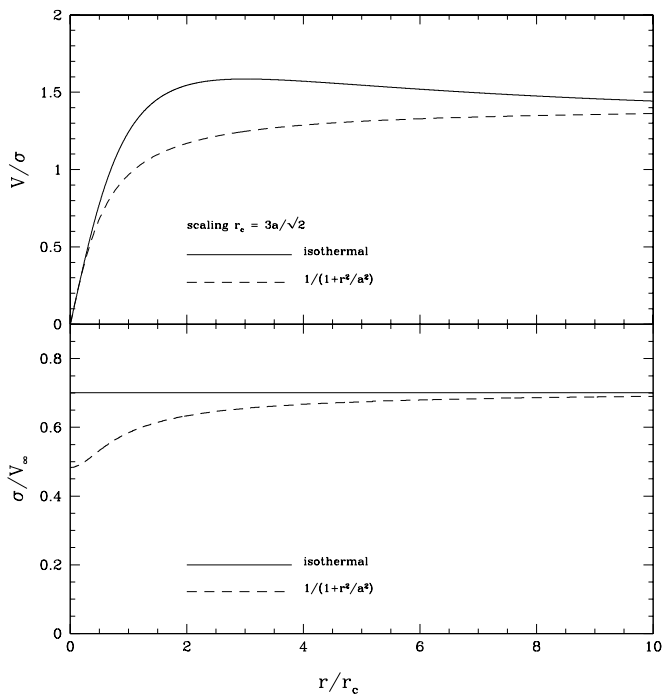} 
  \caption[Rotation curve and velocity dispersion for isothermal and
  pseudo-isothermal halo models]{This figure shows the difference in
    the rotation curve and velocity dispersion of isothermal and
    pseudo-isothermal halo models. The top panel shows that rotation
    of an IT halo rises above the asymptotic rotation speed at large
    radii, before declining to it, while the rotation curve of a PIT
    halo approaches the asymptotic rotation from below. Both rotation
    curves are scaled to the same asymptotic rotation speed to allow
    comparison.  The lower panel shows that an isotropic
    pseudo-isothermal halo is not isothermal, exhibiting a lower
    velocity dispersion at low radii. (Adapted from Fig.~1 of
    Kormendy \&\ Freeman, 2004.)}
  \label{fig:ch7-it_pit}
\end{figure}

Inspection of the $dK_{z,tot}/dz - dK_{z,g}/dz$ difference (cyan; `halo+stars')
shows a steep gradient at small $R$, flattening at large radii,
particularly where the gas layer undergoes exponential flaring at
radii outside 7 kpc.  We see from
Fig.~\ref{fig:ch7-dKdz_halo_vary_q} that this is similar to the
characteristic shape of $dK_{z,h}/dz$ for pseudo-isothermal halos with
different halo flattenings. But despite this similarity, it was
not possible to fit the halo flattening $q$ while holding the core
radius $R_c$ and the asymptotic halo rotation $v_{h,\infty}$ fixed,
even with a zero mass stellar distribution (stellar $M/L_R=0$) and
allowing $q$ to range between oblate and prolate shapes. By adjusting
$q$, and keeping $M/L_R|_{stars}$ small, it is possible to get a similar
shape to the difference gradient (cyan; `halo+stars'), 
but it is always offset to
larger negative values of the gradient. This implies that the
asymptotic halo rotation scale derived from the rotation curve is too
high, as the magnitude of $dK_{z,h}/dz$ is proportional to the
$v_{h,\infty}^2$ \citep[see ][]{srjf1994}.

This may be an artefact of the adopted pseudo-isothermal halo model.
While the flattened pseudo-isothermal (PIT) model is computationally
convenient, we note that a true (spherical) non-singular isothermal
(IT) model was initially adopted for dark halos 
(Carignan \&\ Freeman 1985, 1988).
Few studies have compared the relative merits of the PIT and IT
dark halo models. In their paper on halo scaling laws for disk
galaxies (Sc and later) and dwarf speroidals, \citet{kf2004} compare
halo fits to rotation curves over a large sample and generate scaling
laws between halo parameters measured with a IT halo and those with
a PIT halo. In Fig.~\ref{fig:ch7-it_pit} (adapted from Fig.~1 of
Kormendy \&\ Freeman, 2004), we show IT and PIT halo rotation scaled
to the same asymptotic rotation. As can be seen, the rotation of
the IT halo rises above the asymptotic rotation speed before declining
to it at large radii, while the rotation curve of a PIT halo
approaches the asymptotic rotation from below. The declining shape
of the rotation curve for the IT model would provide a lower and
possibly more realistic estimate of the asymptotic rotational
velocity $v_{h,\infty}$ from a rotation curve decomposition of
rotation data which in practice does not extend to radii in excess
of a few halo core radii.

If an IT halo was fitted to the observed rotation curve
of UGC7321, the asymptotic rotation would be approximately 20-40\%\
lower than that of the PIT halo. This would provide the lower
asymptotic rotation scale $v_{h,\infty}$ necessary to fit the
difference $dK_z/dz$ of UGC7321 with a flattened halo over radial
ranges from 1.5 to 9 kpc.

Flattened non-singular isothermal halos could be formed by halo
rotation or anisotropy of the velocity dispersion.  The rotation
is unlikely to be figure rotation, as figure rotation of triaxial
halos measured in n-body simulations was found to be very slow
\citep{bs2004} ($0.148\,h$ \kms\ kpc$^{-1}$, where $h$ is $H_0/100$,
insufficient to flatten halos more than $q\sim0.7$). The velocity dispersion anisotropy
of the halo dark matter would allow either prolate or oblate halos, just as 
velocity anisotropy of the stars in the brighter elliptical galaxies defines 
the galaxy shape.

With the asymptotic rotation as a free parameter in addition to the
stellar $M/L_R$ and $q$, we found that the residual $dK_z/dz$ curve
(`halo+stars', cyan) is
best modelled with a halo shape of $q=1.0\pm0.1$. Robust least squares
minimization fitting using a Levenberg-Marquardt algorithm (MINPACK-1)
favoured a zero mass stellar disk, but fits were almost as good for an
$M/L_R=0.2$ stellar disk. These fits were successful over the radial
range from $2-9$ kpc. 

\begin{figure*}[ht]
  \centering
\includegraphics[width=15cm]{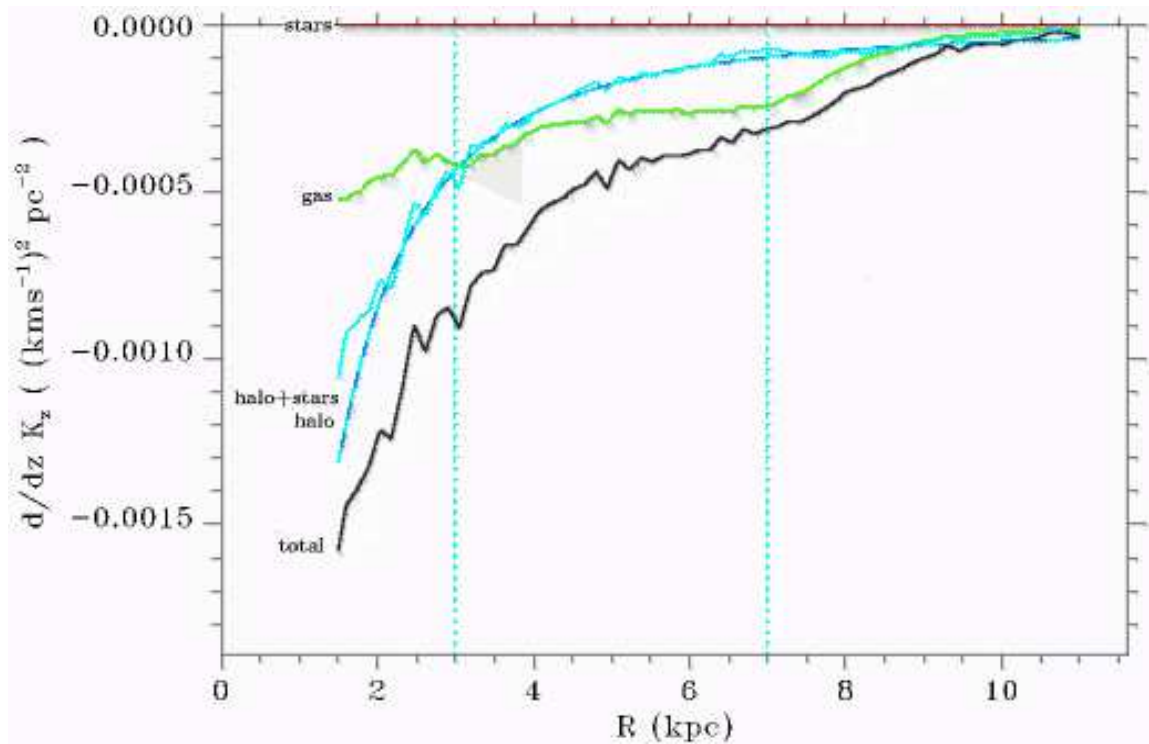}
  \caption[UGC7321: Halo fits to the $z$-gradient of the total
  vertical force for stellar $M/L_R=0$]{The $z$-gradient
    of the vertical force showing successful fits of the halo shape
    for a zero mass disk $M/L_R=0$. The black curve, labelled `total' 
shows the total gradient
    $dK_{z,tot}/dz$ derived from hydrostatics. The green (`gas') and red
(coincident with the abcissa at the top of the figure)  curves show 
the $z$-gradient derived using Poisson's equation for
    the gas and stellar mass components. The
    cyan (`halo+stars') line shows the difference 
$dK_{z,tot}/dz - dK_{z,g}/dz$. This line is
    fitted by the halo gradient $dK_{z,h}/dz$ which is
    modelled by a flattened pseudo-isothermal halo. To achieve a
    successful fit to the cyan line, the asymptotic halo rotation was
    scaled down by 30\%.  The measured halo flattening was $q=1.0\pm0.1$. 
The vertical 
    dotted (cyan) lines show the radial regime used for the fit.}
  \label{fig:ch7-q_fit9}
\end{figure*}

We illustrate this first for the unphysical case where there is no mass in stars 
in Fig.~\ref{fig:ch7-q_fit9}. This figure
is organised in the same manner as Fig.~\ref{fig:ch7-grad_mleq1}.
The gradient due to the stars is now zero at all $z$. Recall that
the (cyan) line `halo+stars' is the observed gradient, which has
to be fit. The smooth (also cyan) line `halo' is that fit (the
dashed -blue- line superimposed is that of the halo alone, which
is the same when $M/L_R$ is zero). This best fit was achieved with
an asymptotic PIT halo rotation reduced by $30\pm5\%$ compared to
the PIT fit to the rotation curve in Fig.~\ref{fig:ch7-rc_fit}. 

It is remarkable that the shape of the $R$-dependence of $dK_z/dz$
for the adopted halo model in Fig.~\ref{fig:ch7-q_fit9} agrees so well with the
shape of the $K_z$ gradient derived from the \HI\ flaring and velocity
dispersion, at least for radii $< 9$ kpc. Although some rescaling of
the {\it strength} of the $K_z$ force was needed, we see that the
density distribution of the adopted spherical PIT, using the core
radius derived from the rotation curve fit, also provides the correct
radial variation of the $K_z$ gradient.  This need not have happened.
Although the $K_R$ estimate from the rotation curve and the hydrostatic
estimate of $dK_z/dz$ come from analysing the same XV data (see paper
III), the two functions come from different features in the XV data, so
are relatively independent.

%%% Figure %%%
% dK_z/dz fits

%%% Figure %%%
% dK_z/dz fits
\begin{figure*}[t]
  \centering
\includegraphics[width=15cm]{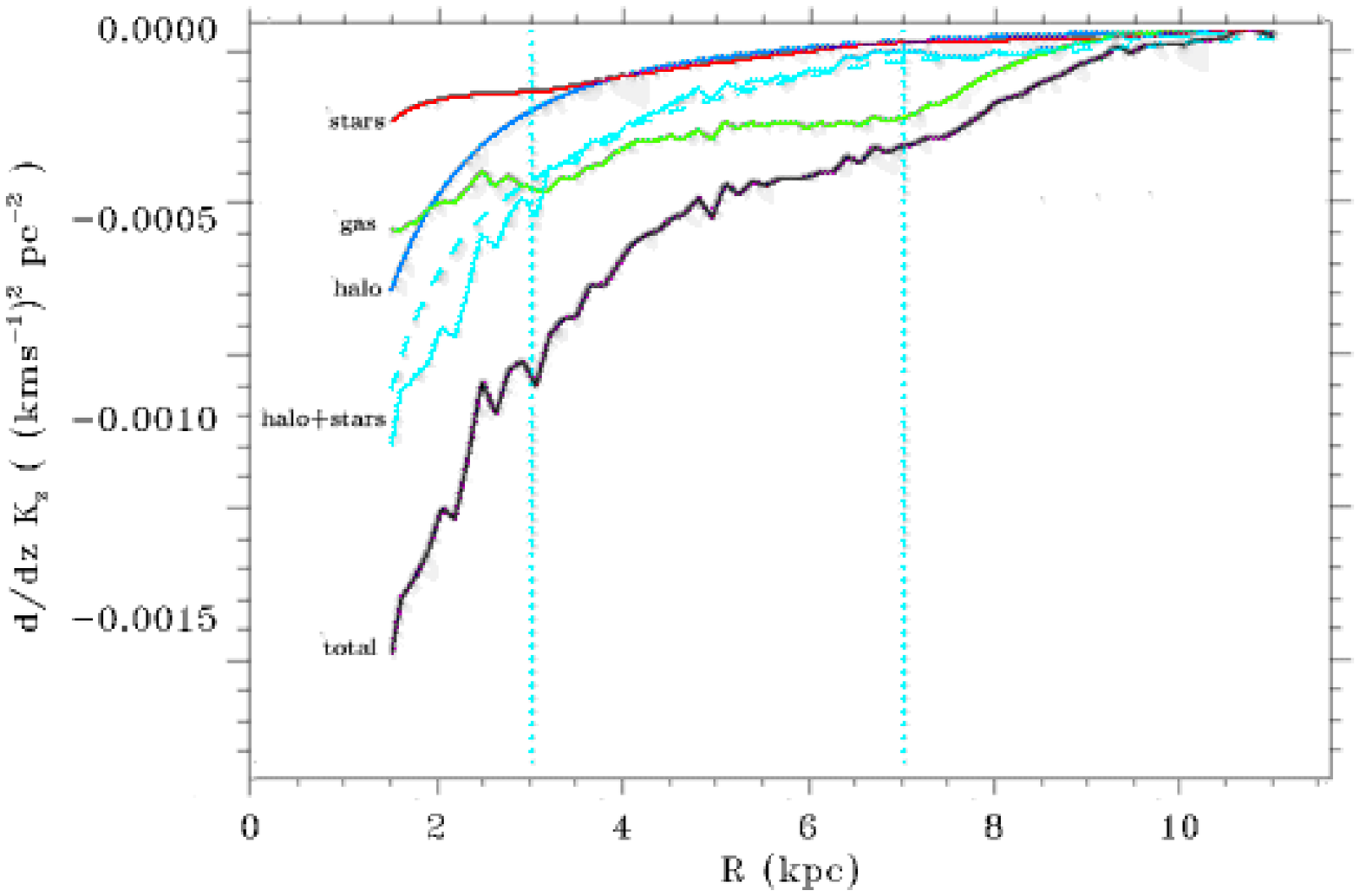}
  \caption[UGC7321: Halo fits to the $z$-gradient of the total
  vertical force for stellar $M/L_R=0.2$]{The $z$-gradient 
    of the vertical force showing successful fits of the halo shape
    for stellar mass-to-light ratio $M/L_R=0.2$. The curve are coloured 
    as in Fig.~\ref{fig:ch7-q_fit9}, except that the line for the
    stars is now at non-zero values. The
    cyan line, labelled `halo+stars', shows again the difference 
$dK_{z,tot}/dz - dK_{z,g}/dz$ which
    is fitted by the stellar gradient $dK_{z,s}/dz$ derived from
    Poisson's equation and the halo gradient $dK_{z,h}/dz$ which is
    modelled by a flattened pseudo-isothermal halo. To achieve a
    successful fit to this line, the asymptotic halo rotation was
    scaled down by 50\%. The thin full-drawn (blue) line, labelled `halo'
corresponds to this halo, while the dashed (cyan) line shows the
    gradient due to this halo and that of the stars together.
A stellar mass-to-light ratio
    significantly larger than 0.2 was not possible. Again,
    the measured halo flattening was $q=1.0\pm0.1$. The vertical
    dotted (cyan) lines show the radial regime used for each fit.}
  \label{fig:ch7-q_fit10}
\end{figure*}

In  Fig.~\ref{fig:ch7-q_fit10} we show the fit for a stellar $M/L_R=0.2$ disk. 
The lines labelled `halo+stars' (cyan) show the gradient as deduced from the
observed total gradient minus that of the gas in the full-drawn line (which of
course is the same as in Fig.~\ref{fig:ch7-q_fit9}) and that of 
the sum of the gradient of the halo model
fit and that deduced from the stellar distribution with $M/L_R=0.2$ as
the dashed line. The gradient from the stellar disk alone is the (red) curve
labelled `stars' and for the halo the (blue) curve labelled `halo'. 
For this case an asymptotic halo rotation reduction of $50\pm5\%$ was needed. 
In effect, reducing $v_{h,\infty}$
and reducing the stellar $M/L_R$ have similar effects of increasing the
magnitude of the asymptotic value of the difference $dK_z/dz$ curve
(`halo+stars',cyan). For
both cases ($M/L_R=0$ in  Fig.~\ref{fig:ch7-q_fit9} and $M/L_R=0.2$ in
Fig.~\ref{fig:ch7-q_fit10}), the shape of this difference $dK_z/dz$ curve 
dictated (given the derived core radius for the dark halo from our
rotaton curve decomposition) a halo flattening close to spherical.

At radii larger than 9 kpc, the strong flaring causes the difference
$dK_z/dz$ (cyan, `halo+stars') to be too small to be fit with the 
same asymptotic halo rotation. 
Even the shallow gradient of $dK_z/dz$ given by a highly
prolate halo, combined with a low $v_{h,\infty}$, did not produce a
good fit at these radii. We briefly discuss why the derived gradient for
halo+stars may have been underestimated.

We have argued that a true isothermal halo may provide a more valid
model for this analysis. Another possibility is that the gas velocity
dispersion is not vertically isothermal. We were forced to adopted this
assumption in the hydrostatic equation, because there is currently no available
measurements of the $z$-dependence of the gas velocity 
dispersion.\footnote{Because of S/N limitations, our measurement of the 
gas velocity dispersion in paper III models the H{\scriptsize I}
XV diagram integrated over $z$. Thus it represents a luminosity-weighted
average dispersion as a function of radius.} Prior to our work the
gas velocity dispersion had only been measured in a few face-on
galaxies. Our \HI\ disk modelling of edge-on galaxies has more than
doubled the number of galaxies with radial gas velocity dispersion
measurements. These high resolution observations show that the gas
velocity dispersion is not isothermal in radius, but its vertical
properties are unknown and we had to assume that is is vertically
isothermal.

The gas velocity dispersion in disk galaxies is often ascribed to local 
heating by supernovae and stellar winds in star formation regions. 
Indeed, \citet{svdk1984} found in NGC628 that the gas velocity dispersion
is systematically higher in the spiral arms than in between. 
On the other hand, similar velocity dispersions are seen in
regions of star formation and in regions where there is no visible
star formation \citep[e.g. ][]{mcbf1996}, and both low and high
surface brightness galaxies seem to have similar gas velocity
dispersion.  \citet[][]{sb1999} offer a plausible alternative,
suggesting that weak magnetic fields in galaxies allow energy to be
extracted from differential rotation via MHD-driven turbulence. This
would result in a gas velocity dispersion that was proportional to the
rotational shear due to the disk, resulting in similar gas velocity
dispersion for galaxies with similar rotation curves.  
However, while heating caused by gas shear
could generate radial variation in the gas velocity dispersion, it is
unclear how it could cause the gas velocity dispersion not to be
vertically isothermal. Conversely, the decline in star
formation away from the midplane could cause a fall-off in gas
velocity dispersion with $z$.

A non-isothermal vertical gas velocity dispersion would
probably have more of an effect at larger radii where the flaring
gas is probing a larger range in $z$. A gas velocity dispersion
declining with $z$ would increase the absolute total vertical
gradient of $K_z$ derived from the equation of hydrostatic equilibrium.
A significant increase of $dK_{z,tot}/dz$ would enable a
larger asymptotic halo rotation more consistent with a pseudo-isothermal
halo and a larger stellar $M/L_R$.

Another, less plausible, explanation of the $dK_z/dz$ fitting problem at
large radius, is that the gas-to-\HI\ ratio used to scale the \HI\ density
to account for \He\ and \Htwo\ is not constant. This is unlikely as the
\He\ content is mainly primordial and well known from big bang
nucleo-synthesis.  As \He\ accounts for 0.34 of the additional
0.4 fraction, it is unlikely that a radially declining molecular
hydrogen distribution could significantly reduce $dK_{z,g}/dz$ thus
allowing a higher difference $dK_z/dz$ (cyan).

\section{Comparison to other work}
\label{sec:ch7-comparison}

We first review earlier work on the flattening of dark
halos in spiral galaxies. The earliest concern was whether 
the dark matter indicated by flat rotation curves resided indeed in 
a more or less round halo or was part of the disk. 
That the latter was not the case was
shown in 1981, using evidence from bulge isophotes in external
galaxies and star counts in our Galaxy \citep{mrs81} and from \HI\ flaring
in NGC 891 \citep{vdkruit1981}. Next, the question of the actual
flattening $q = c/a$ of dark
halos in spiral galaxies arose and we will now review previous work on this
subject, starting with our Galaxy.  One of the
early methods is the analysis of the local surface density in the
Solar neighbourhood using stellar kinematics. 
With this method \citet{bmo1987} find $0.3 \leq q \leq
0.6$, \citet{vdmarel1991} $q \geq 0.34$ and  \citet{bsmks2005} $q
\gesim 0.5$.

At large radial distances of $8.0 \leq R \leq 60$ kpc, RR Lyrae stars
show the dark matter distribution to be flattened by $q \sim 0.7$
\citep{ac1994}. Hyper-velocity stars open another promising way of
probing the shape of the Galactic dark matter distribution. One star,
assuming it is $70$ kpc away, gives $0.5 < q < 1.6$  \citep{ggmez2005}. 
\citet{scmz1999} used the microlensing optical depth towards the
Galactic bulge, LMC, SMC and M31 to probe the shape of the Galactic halo to
large radii ($R\leq 5 R_{25}$). However, they were not able to derive 
strong constraints: $q =0.6 \pm 0.4$.

Since the discovery of the Sagittarius dwarf galaxy, modelling of its
extended stellar tidal debris stream has become one of the most
promising methods. \citet{mswo2003} show that the Sagittarius stream traces a
great circle around our Galaxy, extending to radii of $2 R_{25}$ from
the Galactic centre. If the tidal debris has made several orbits, the
Galactic halo must be near-spherical so that the stream does not precess away
from a single plane. \citet{merrifield2004}
argued that the apparent coherency of the carbon star kinematics in
the stream suggest that all the stars are on the same wrap, making it
impossible to constrain the halo flattening. Conversely,
\citet{ilitq2001} contend that the stream has made several orbits, and from
this infer that the Galactic halo must have flattening $q \gesim 0.7$
in the radial range $16 < R < 60$ kpc.

Recently, numerically modelling of small satellite infall on a
Sgr-like orbit by \citet{helmi2004b} finds that tidal streams younger
than about $2$ Gyr lead to spatially coherent streams for a large
range of halo flattenings $0.6 \leq q \leq 1.6$. Since then she
\citep{helmi2004c} has significantly revised her initial measurement
to a highly prolate shape with $1.25 \lesim q \lesim 1.5$ by
constraining the star sample to the older Sgr stream stars of
\citet{lmsj2004}. However, \citet{jlm2005} dispute this result,
finding a near-spherical halo with $q \sim 0.83-0.92$.  In a more
recent analysis of the Magellanic Stream \citet{rpt2007} find a
flattening of $0.74\le q \le 1.20$.

The situation for halo shape measurement in external galaxies is just
as confusing, because some methods are suited only to specific types of
galaxies. The determination of halo shape from polar ring galaxies is
such a case. By simply comparing the
equatorial and polar rotation curves it is possible to ascertain the
flattening of the total potential. Using
this method NGC4650A and A0136-0801 were found to be moderately
flattened with $q \gesim 0.7$
\citep[][respectively]{wmes1987,swr1983}, while MCG-5-7-1 was found to
be approximately spherical \citep{wmes1987}. 
A potentially more accurate method is to model the rotation along both
axes using a multi component mass model comprising bulge, equatorial
stellar and gas disks, and polar stellar and gas rings. Using this
method, \citet{ss1990} originally found the halo flattening of
NGC4650A to be $0.3 \leq q \leq 0.7$; subsequent higher quality
observations were able to constrain the halo more tightly, to $0.3
\leq q \leq 0.4$ \citep{srjf1994}. This method has also been
applied to AM2020-504, where the flattening was found to be
$q \sim 0.6$ \citep{acchs1993}.
Another method involves modelling of the twisting caused by precession 
of the ring. With some specific assumptions, 
\citet{sckd1992} constrain the flattening of the NGC4753 halo to be
$0.84 \leq q \leq 0.99$.
Finally, using the twisting
of the morphological minor axis of the disk plane away
from the kinematic minor axis to model the velocity field of polar
rings, the flattening of the dark halo of A0136-0801 was found to be
$q \sim 0.6$ \citep{sp1995}.

Another method that has been used to measure halo flattening is strong
gravitational lensing. An early study of a double lens system
comprising two spirals found $q \gesim 0.4$ \citep{kdbj1998}. More
recently there have been two studies of multiple quad lens systems
finding $q \gesim 0.4$ \citep{rt2001} and $q \sim 0.7$ \citep{ck2004},
and another analysis of a double lens system $0.6 < q < 0.7$
\citep{chaeetal2002}.

Warps in stellar disks \citep[e.g.][]{rc1998} 
offer several mechanisms to probe the halo
shape of spiral galaxies. One method uses 
the precession of the warped disk to
constrain the halo flattening. It has been
applied to NGC2903, yielding a halo flattening of $q =
0.80 \pm 0.15$ \citep{hs1994}.

It is also possible to measure the mean shape of vast numbers of
galaxies via weak gravitational lensing. 
Measurements of about $10^5$ lensed systems against about $10^6$ background
galaxies  \citep{hyg2004a} find a
mean projected halo ellipticity of $0.20^{+0.04}_{-0.05}$ and a mean
projected halo flattening of $\langle q \rangle =
0.66_{-0.06}^{+0.04}$ (1-$\sigma$ error) .  However, a larger
investigation of about $2$ million lensed galaxies  against 
$32$ million background galaxies from the SDSS dataset found no strong
evidence of flattening, with $\langle q \rangle = 0.99_{-0.05}^{+0.06}$
\citep{mhbsb2005}.

The results of halo flattening studies so far do not reveal a
consistent picture.  We believe that the method of the flaring of
the gas layer is among the most promising, at least for late-type
spiral galaxies.   First tried by \citet{crb1979} on the Galaxy,
early development of the method was undertaken by \citet{vdkruit1981}
who applied it to low resolution observations of NGC891, concluding
that the halo was not as flattened as the stellar disk. It was then
applied to several galaxies in the 1990's, most notably the careful
study of the very nearby Sc galaxy NGC4244, which found a highly
flattened halo with by $q=0.2^{+0.3}_{-0.1}$ out to radii of $\approx
2 R_{25}$ \citep{olling1996b}. All applications of the flaring
method have indicated  highly flattened halo distributions with $q
\leq 0.5$ \citep{becquaert1997,bc1997a,sicking1997}.  Recently,
\citet{bj2008} measured a flattening of $q=0.4$ from flaring of the
\HI\ layer in M31. This assumed a constant \HI\ velocity dispersion
with radius; if it is allowed to have a modest decline in the outer
disk the flattening can be made less with $q$ more like 0.5 to 0.6.
With the exclusion of NGC4244, it may therefore be suspected that
the assumption of a radially constant gas velocity dispersion has led
to errors in the derived flattening of the halo.

Measurements of significant flattening using the flaring method
initially led to the supposition that perhaps the method is
systematically biased to flattened halos. Our analysis of UGC7321
shows that this is not the case: the gas layer flaring method is
just as sensitive to prolate halos as it is to oblate ones. Here,
we briefly consider the set of $q$ measurements using the flaring
method that have indicated flat halos. The flattening for NGC891 
\citep{bc1997a} was estimated from 
VLA observations with a low peak signal-to-noise of 13
\citep{rupen1991}. The low sensitivity could have led to underestimates
of the gas density and vertical flaring, thus changing the shape of
both $dK_{z,tot}/dz$ and $dK_{z,g}/dz$, and thereby $q$. Except in
the case of NGC4244 \citep{olling1996b}, it is unclear what model
was used for the radial gas velocity dispersion. An assumption of
radially-constant gas velocity dispersion could easily skew the
derived halo shape measurement.

In some cases it is
not clear whether the gas self-gravity was included in the mass modelling.
Additionally, excluding NGC4244, all the previous measurements of the
halo flattening from the gas layer flaring were performed on large Sb-Sc
galaxies with maximum rotation speeds $v_{max}$ between 177 and
295 \kms. As the gas layer flaring is inversely proportional to
$v_{max}$, the maximum \HI\ flaring of these galaxies is $\lesim 1$
kpc, making it difficult to resolve unless the galaxy is
nearer than 5 Mpc.

The Galactic $q$ measurement from the gas flaring by \citet{om2000} is
particularly interesting. They were unable to fit the
halo with a pseudo-isothermal model, unless the Solar radius and
rotation velocity are significantly less ($R_{\odot}\leq7.6$ kpc,
$\Theta_{\odot}\leq190$ \kms) than the standard values. 
The uncertainty associated with these values translates to a large
uncertainty of $q$: $q=0.8\pm0.3$.

UGC7321 is the least massive galaxy for which the halo flattening has
been measured. The derived $R$-band face-on central surface brightness
is 2.5 times fainter than the $B$-band measurement of NGC4244, and the
total $R$-band luminosity is 4.5 times fainter than NGC4244, while its
gas layer flares to twice the height of NGC4244. The very low stellar
mass of UGC7321 made it an ideal candidate for halo
modelling with the gas flaring method. 

Although there are now a number of different measurements of galactic
halo flattening, there is no obvious concentration around a particular
halo shape or any correlation of halo flattening with galaxy
morphology. Currently the measured $q$ values range from 0.1 to
1.4.  The low $q$ values for the large Sb galaxies, M31,
NGC891 and NGC4013, are puzzling as in these cases the stellar
density distribution may be more spherical than the halo density
distribution. It seems
unlikely that the galactic halos could exist in the range of shapes
measured, unless the fractions of the constituent dark matter types
vary significantly from galaxy to galaxy. Early work by
\citet{dubinski1994} found that including baryon infall in n-body
halo simulations led to nearly axisymmetric halos. Most n-body
simulations without hydrodynamics tend to form prolate halos
\citep{sellwood2004}; however, new work by Dubinski (unpublished)
has shown that the inclusion of hydrodynamical modelling generates
halos that are more spherical.

We note here an application of \HI\ hydrostatics to our Galaxy by
\citet{kdkh07}, which illustrates the potential power of
\HI\ hydrostatics to trace the Galactic potential gradient and hence
the total dark matter distribution in the Galaxy.  Kalberla et al.
adopted an isothermal velocity dispersion for the Galactic \HI\,
and found several components of dark matter, including the usual
extended halo with a mass of about $1.8 \times 10^{12}$ \msun, a
thick self-gravitating disk with a mass of about $2$ to $3 \times
10^{11}$ \msun, and an outer dark matter ring with a mass of about
$2$ to $3 \times 10^{10}$ \msun.  Similar studies in other edge-on
galaxies may reveal comparable substructure in the dark matter,
including dark matter rings which may be left over from accreted
and circularized smaller galaxies, drawn down into the disk by
dynamical friction.  As we will argue in the next section, it is 
important to measure the structure, rotation and velocity
dispersion of the \HI\ in both $R$ and $z$, to ensure such structures
are not artifacts of assumptions required to apply the hydrostatics.

Finally we note that since we submitted the original version of this paper,
a study of the density distribution of dark matter halo of UGC 7321 by \citet{bmj10} 
appeared, using the rotation curve and flaring of the \HI\ layer derived from the same 
data. In this study the fitting was performed with the halo central density, 
core radius and radial exponential density slope as free parameters, but with the
halo assumed spherical, the stellar $M/L$ of the disk fixed and using values for the 
\HI\ velocity dispersion from Gaussian fits to the position-velocity profiles (typically
7 to 9 \kms). These authors also conclude from their work
that the dark matter halo dominates the dynamics of UGC 7321 at all
radii, but they do not put constraints on the dark matter halo flattening.

\section{Conclusions}
\label{sec:ch7-summ}

In this study we have shown that it is possible to measure the gas
flaring and \HI\ velocity dispersion via modelling of the \HI\
distribution. Using these methods we found that the small late-type
disk galaxies in our sample show substantial \HI\ flaring, increasing
linearly with radius in the inner disk and exponentially in the
outer disk. The \HI\ velocity dispersion has a mean value of 7 \kms,
but varies from 4.5 to 12 \kms. Our \HI\ modelling method is also
capable of measuring the vertical variation of the \HI\ velocity
dispersion given additional \HI\ observations.

UGC7321, a small low surface brightness Sd galaxy in our sample,
has the most accurate flaring measurements in our sample. We were
unable to model the observations using a pseudo-isothermal halo.
By lowering the asymptotic halo rotation to a value corresponding
to a true isothermal halo model, we found that UGC7321 has a spherical
halo density distribution of $q=1.0\pm0.1$.  Highly prolate halos
($q>1.2$) and highly flattened halos ($q<0.6$) are strongly excluded
if our approximation of a true isothermal halo is valid.

Our mass modelling analysis assumed that the \HI\ gas velocity
dispersion was vertically isothermal, as no measurements of the
vertical variation of the \HI\ gas velocity dispersion are as yet
available.\footnote{Our measurements of the \HI\ gas velocity
dispersion used the vertically averaged \HI\ distribution, ie. they
are the luminosity-weighted mean velocity dispersion as a function
of radius.} If the \HI\ velocity dispersion is in fact vertically
declining, this would lead to a larger estimated vertical gradient
of the total vertical force, which may allow a pseudo-isothermal
model for the halo.

UGC7321 is a gas-rich galaxy ($M_{\HI}/L_R=2.2$), with a very low stellar
mass galaxy ($M=3\times10^8$ \msun), four times less massive than the
gas disk. The $R$-band stellar mass-to-light ratio of UGC7321 is very
low at $M/L_R\lesim0.2$. Mass modelling of the vertical force
distribution showed that vertical force fitting provides a much
stronger constraint on the stellar mass-to-light ratio than the
standard method of radial force fitting via rotation curve
decomposition.

Two important assumptions in this work need to be tested further.
The first is that the \HI\ velocity dispersion is isothermal in
$z$. For a definitive estimate of the $R$ and $z$-components of the
total potential gradients from \HI\ hydrostatics, it is essential
to have reliable measurements of the \HI\ density, rotation and
velocity dispersion as a function of both $R$ and $z$.  It should
be possible, with additional short spacing ATCA observations
supplementing our data, to measure the \HI\ velocity dispersion as
a function of $z$ in ESO274-G001 by modelling the \HI\ XV diagram
at varying heights above the galactic plane.  ESO274-G001 is the
closest, isolated, southern edge-on galaxy at a distance of 3.4
Mpc.  In the northern hemisphere, UGC7321 is a prime candidate due
to its high \HI\ mass, despite its larger distance of 10 Mpc.  The
large \HI\ flaring means that the \HI\ could be measured at a height
of 400 pc for radii from 5-11 kpc, and at 700 pc for radii from
9-11 kpc.  The dwarf Scd galaxy NGC5023 is also an excellent
candidate, given its distance of about 8 Mpc \citep{vdks1982a}.
For this galaxy, early flaring measurements by \citet{bsvdk1986}
found that the gas thickness was constant with radius. This is a
surprising result, because the $v_{max}$ value for this galaxy is
only about $80$ \kms, and large flaring might be expected.  It would
be interesting to measure the radial and vertical variation of the
gas velocity dispersion, gas flaring, and halo shape using beter
data, as this galaxy has a similar size, \HI\ brightness and total
mass as UGC7321.

The other important test is to determine whether a true isothermal
halo provides a better model than the pseudo-isothermal halo for
the dark matter in late-type disk galaxies, or whether there are
better models than either of these.  Our analysis of UGC7321
has shown that the vertical gradient of the vertical force provides
a significantly stronger constraint on the halo density distribution
than does rotation curve decomposition. So, this test can in principle
be achieved by analysing UGC7321 and the other galaxies in
our sample with both flattened pseudo-isothermal and true isothermal
halo models. Such flattened isothermal halos could be flattened by
rotation or by anisotropy of the velocity dispersion.  This will
determine which kind of model is better for both the radial halo
force as measured from the rotation curve, and the vertical force
of the halo determined from $dK_z/dz$ fitting.

%%% Local Variables: 
%%% mode: latex
%%% TeX-master: "thesis"
%%% End: 

\begin{acknowledgements}

We are very grateful to Albert Bosma who contributed greatly to
initiating this project. He pointed out that \HI\ flaring studies are
best done on edge-on galaxies with low maximum rotational velocities,
and we used an unpublished Parkes \HI\ survey of edge-on galaxies by
Bosma and KCF when selecting our galaxies.  JCO thanks E. Athanassoula,
M.  Bureau, R.  Olling, A. Petric and J. van Gorkom for helpful
discussions.  JCO is grateful to B. Koribalski, R. Sault, L.
Staveley-Smith and R.  Wark for help and advice with data reduction and
analysis, and to P. Sackett, A. Kalnajs and F. Briggs for advice and
discussions on the modelling.  M. Pohlen generously provided his deep
$R$-band image of UCG7321, for which we are very grateful. Scott Tremaine 
has our gratitude for a few important comments on a draft version of this paper.
We thank the referee, J.M. van der Hulst, for
his careful and  thorough reading of the manuscripts of this series of papers and
his helpful and constructive remarks and suggestions.

\end{acknowledgements}

\bibliographystyle{aa}

\end{document}